\documentclass[twocolumn,aps,superscriptaddress,multicol,amsmath,amssymb]{revtex4-2}
\usepackage{amssymb}
\usepackage{amsmath}
\usepackage{graphicx}
\usepackage{sidecap}
\usepackage{epstopdf}
\usepackage{bm}%
\usepackage{gensymb}
\usepackage{soul}
\usepackage{sidecap}
\usepackage{bbm}
\usepackage[normalem]{ulem}
\UseRawInputEncoding

\newcommand{\be}{\begin{equation}}
\newcommand{\ee}{\end{equation}}
\newcommand{\bk}{{{\bf{k}}}}
\newcommand{\bq}{{{\bf{q}}}}

\newcommand{\br}{{{\bf{r}}}}

\newcommand{\bea}{\begin{eqnarray}}
	\newcommand{\eea}{\end{eqnarray}}

\newcommand{\bd}{\begin{displaymath}}
	\newcommand{\ed}{\end{displaymath}}
\newcommand{\ba}{\begin{array}}
	\newcommand{\ea}{\end{array}}
\newcommand{\bi}{\begin{itemize}}
	\newcommand{\ei}{\end{itemize}}
\newcommand{\bc}{\begin{center}}
	\newcommand{\ec}{\end{center}}
\newcommand{\bfl}{\begin{flushleft}}
	\newcommand{\efl}{\end{flushleft}}
\newcommand{\bfr}{\begin{flushright}}
	\newcommand{\efr}{\end{flushright}}

\newcommand{\mi}{\rm i}

\newcommand{\bl}{\begin{aligned}}
	\newcommand{\el}{\end{aligned}}

\def\br{{\bf r}}
\def\bk{{\bf k}} \def\bq{{\bf q}}  
  \def\bd{{\bf d}}

\def\6{\partial}

\def\o{\omega}

\def\={\!\!\!&=&\!\!\!}
\def\+{\!\!\!&&\!\!\!+~}
\def\-{\!\!\!&&\!\!\!-~}

\newcommand\redout{\bgroup\markoverwith{\textcolor{red}{\rule[.5ex]{2pt}{0.4pt}}}\ULon}

\usepackage[colorlinks=true,citecolor=blue]{hyperref}
\hypersetup{colorlinks=true,citecolor=blue,linkcolor=red,urlcolor=blue}

\newcommand{\orcid}[1]{\href{https://orcid.org/#1}{\includegraphics[width=8pt]{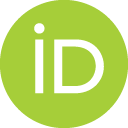}}}

\graphicspath{{.}{./Figs/}}

\newcommand\bluesout{\bgroup\markoverwith{\textcolor{blue}{\rule[0.5ex]{2pt}{0.4pt}}}\ULon}
\usepackage{hyperref,cleveref}
\begin{document}
\title{
Topological Superconductivity in Sn/Si(111) driven by non-local Coulomb interactions}
%
\author{Mehdi Biderang \orcid{0000-0002-6666-1659}} 
\email{mehdi.biderang@umanitoba.ca}
\affiliation{Department of Physics and Astronomy, University of Manitoba, Winnipeg R3T 2N2, Canada}
\affiliation{Manitoba Quantum Institute, University of Manitoba, Winnipeg R3T 2N2, Canada}
\author{Mohammad-Hossein Zare \orcid{0000-0003-4638-7987}}
\affiliation{Department of Physics, Qom University of Technology, Qom 37181-46645, Iran}
\author{Jesko Sirker}
\email{sirker@physics.umanitoba.ca}
\affiliation{Department of Physics and Astronomy, University of Manitoba, Winnipeg R3T 2N2, Canada}
\affiliation{Manitoba Quantum Institute, University of Manitoba, Winnipeg R3T 2N2, Canada}
\date{\today}
%
%
\begin{abstract}
	Superconductivity was recently observed in boron-doped ($\sqrt{3}\times\sqrt{3}$)Sn/Si(111). The material can be described by an extended Hubbard model on a triangular lattice. Here, we use the random-phase approximation to investigate the charge and spin fluctuations as well as the superconducting properties of the system with respect to filling and the relative strength of the extended versus the on-site Hubbard interactions.
	Our calculations reveal that near half-filling and weak extended Hubbard interactions, the superconducting ground state exhibits chiral $d$-wave pairing.
	Far from half-filling and for stronger nearest-neighbor Coulomb interactions, the system shows chiral $p$-wave (hole-doping) and $f$-wave (electron-doping) pairings.
	The dependence of the pairing symmetry on the extended Hubbard interactions suggests that charge fluctuations play an important role in the formation of Cooper pairs.
	Finally, the temperature dependence of the Knight shift is calculated for all observed superconducting textures and put forward as an experimental method to examine the symmetry of the superconducting gap function.
\end{abstract}
%
\maketitle
%

\section{Introduction}
Unconventional superconductivity, i.e.~superconductivity not described by BCS theory, has been one of the most challenging topics in condensed matter for the last 35 years~\cite{Kamahira_ACS_2008,Biderang_PRB_2017,Greco_PRL_2018,Romer_PRB_2021}.
The discovery of high-$T^{}_{\rm C}$ superconductivity in doped cuprates and iron pnictide alloys triggered 
many attempts to find new mechanisms for Cooper pairing beyond the electron-phonon interaction~\cite{Muller_Bednorz_1986,Stewart_REv_Mod_Phys_2011,Sato_Rep_Prog_Phys_2017}.
The majority of materials showing unconventional superconductivity are Mott insulators and do not host itinerant modes in the undoped case.
Here, the strong Coulomb repulsion between the electrons stops the charge carriers from freely moving in the solid.
However, it has been observed that the doping of electron or holes into the compound may lead to the formation of Cooper pairs and the realization of a macroscopic superconducting state.
It has been proposed that the screening of the repulsive Coulomb interaction can cause magnetic or charge fluctuations which then play the role of a glue to pair up the electrons. 
Despite continuing efforts to better investigate and describe high-temperature superconductors, a generally accepted theory remains elusive.
Two-dimensional (2D) atomic layers deposited on  semiconducting substrates are platforms for the realization of  Mott physics and are used in the electronic industry~\cite{Carpinelli_Nature_1996,Carpinelli_PRL_1997,Weitering_PRL_1997,Perez_PRL_2001}.
These materials have a very simple electronic band structure and can be manufactured by adsorption of 1/3 monolayer of group-IV elements such as lead or tin on a heavily hole-doped silicon (111) substrate~\cite{Profeta_PRL_2007}.
%
%
Among every four valence orbitals of the adatoms, one remains unbonded, generating a dangling bond with only one electron~\cite{Wu_PRL_2020}.  
This half-filled electronic state is subject to strong electron-electron interactions that can be modeled by the Hubbard model with a ($\sqrt{3}\times\sqrt{3})R30^{\circ}_{}$ structure~\cite{Li_NatComms_2013,Wu_PRL_2020}.
The triangular lattice Hubbard model is a standard model to study the competition between strong electronic correlations and geometric frustration~\cite{Wolf_PRL_2022,Biderang_PRB_2022,Gneist_arXiv_2020}.
In addition, it has been suggested that non-local Coulomb interactions in Sn/Si(111) are important~\cite{Wolf_PRL_2022}.
In 2D atomic layers, it is possible to tune the hopping integrals, filling coefficients, and Coulomb repulsion parameters using impurities or by depositing  adatoms on different substrates~\cite{Glass_PRL_2015,Hirahara_PRB_2009}.
This allows to realize a wide variety of exotic phenomena such as metal-insulator transitions, various magnetic states, charge density waves, possible quantum spin liquid states, and chiral superconductivity~\cite{Tresca_PRL_2018,Adler_2019,Wu_PRL_2020}.
Recent experiments based on scanning tunneling microscopy (STM) have detected the signature of superconductivity in Sn/Si(111) with $T^{}_{C}=4.7\pm 0.3$ K~\cite{Nakamura_PRB_2018,Tresca_PRL_2018,Wu_PRL_2020,Machida_PRB_2022}.
Theoretical studies have investigated the superconducting gap function in the presence of Rashba spin-orbit coupling (SOC), which is always present due to the lack of spatial inversion symmetry in the heterostructure.
These studies found, on the one hand, the mixing of  spin-singlet and triplet superconductivity~\cite{Nakamura_PRB_2018}, and, on the other hand, pure spin-singlet $s$-wave pairing~\cite{Machida_PRB_2022}; results which are not consistent with each other.  
%
Besides, a recent theoretical investigation using functional renormalization group (FRG) and weak-coupling renormalization group (WCRG) approaches lead to a phase diagram for the superconducting instability of the system~\cite{Wolf_PRL_2022} consisting of chiral $d$-wave, chiral $p$-wave, and odd-parity spin-triplet $f$-wave pairings.
These contradictory results lead us to believe that more work is required to understand the mechanism for the Cooper pairing in this system.
In this paper, we will investigate the influence of the electron-electron correlations and the level of doping on the charge, spin and superconducting instabilities of ($\sqrt{3}\times\sqrt{3}$)Sn/Si(111) using the random phase approximation (RPA).
Our paper is organized as follows: 
In Sec.~\ref{SubSec:Band_DOS}, we will describe the tight-binding model and find its non-interacting band structure and density of states (DOS).
Then in Sec.~\ref{SubSec:Charge_Spin_Kappa}, we introduce the bare and RPA charge and spin susceptibilities to study the charge and magnetic fluctuations of the system.
More specifically, we investigate the effects of filling, on-site and extended Hubbard interactions on the texture of charge and spin fluctuations.
Section~\ref{SubSec:Eff_Int} describes how to determine the effective interaction in both charge and spin channels in the framework of RPA and these results are used to discuss the superconducting instability in Sec.~\ref{SubSec:SC}.
The obtained results for the magnetic and charge fluctuations and the phase diagram of the superconducting state in ($\sqrt{3}\times\sqrt{3}$)Sn/Si(111) are then presented in Sec.~\ref{SubSec:Res_Orders} and Sec.~\ref{SubSec:Res_SC}, respectively. 
Finally, we calculate the temperature dependence of the Knight shift providing a connection of our results to an experimental technique which can be used to detect the symmetry of the superconducting gap function.
The last section is devoted to a short summary and conclusion.
%
%
\begin{figure}[t]
	\begin{center}
		\vspace{0.10cm}
		\hspace{-0.3cm}
		\includegraphics[width=0.99 \linewidth]{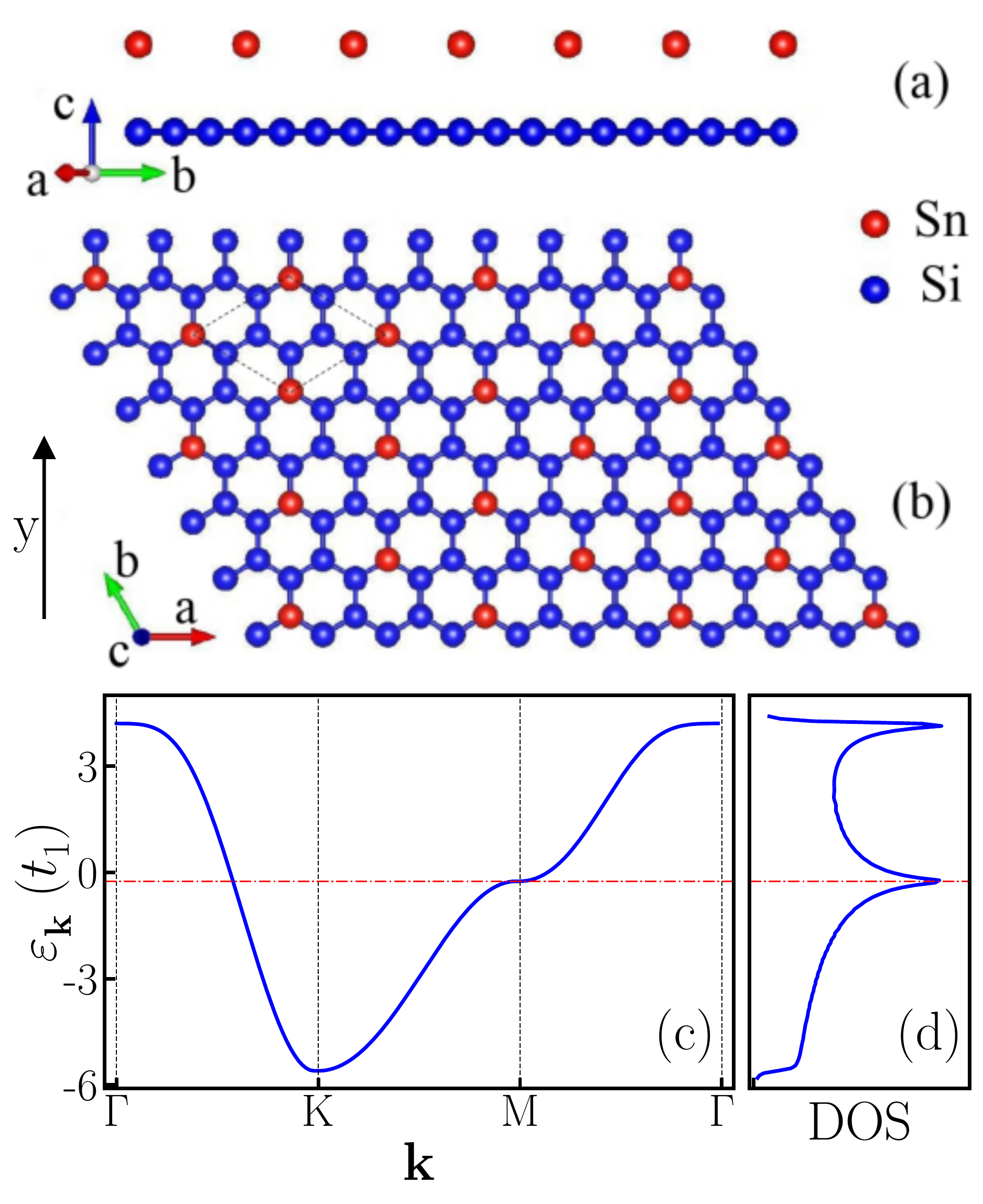}  \hspace{.0cm}
	\end{center}
	\vspace{-0.2cm}
	\caption{
		Schematic representation of the ($\sqrt{3}\times\sqrt{3}$)-Sn/Si(111) system. 
		(a) Side and (b) top views of the effective triangular lattice model which results from the half-filled dangling bonds.
		(c) The obtained band structure along the high-symmetry path,
		and (d) the density of state (DOS).
		Note that the red line crosses the saddle point M, and shows the position of the van Hove singularity at $\langle n \rangle\approx 0.92$.
	}
	\label{Fig:Fig_Geometry_Band}
\end{figure}
%

%
\section{METHODOLOGY}
\label{Sec:Methodology}
In this section, we first review the relevant electronic band structure of Sn/Si(111) following Ref.~\cite{Adler_2019}.
Next, we investigate the spin and charge fluctuations in the system driven by electron-electron interactions using RPA. 
Finally, we obtain the possible superconducting instabilities using the BCS gap equation.

\subsection{Electronic Band Structure of Sn/Si(111)}
\label{SubSec:Band_DOS}
Figs.~\ref{Fig:Fig_Geometry_Band} (a), (b) show the side and top views of a ($\sqrt{3}\times\sqrt{3}$)Sn/Si(111) system, in which the Sn atoms (red) are deposited at $T^{}_{4}$ adsorption sites, above the Si atoms (blue).
The Sn atoms form a ($\sqrt{3}\times\sqrt{3}$) $R30^{\circ}_{}$ superstructure with respect to the ($1\times1$) periodicity of the Si(111) surface.
Therefore, this system is a platform to study electronic correlations on a triangular lattice.
In this system, each Sn atom possesses a dangling bond pointing towards the c-direction and containing only one electron~\cite{Li_NatComms_2013,Wu_PRL_2020}.
In addition, due to the lack of spatial inversion symmetry, an antisymmetric Rashba spin-orbit coupling (SOC) is induced~\cite{Nakamura_PRB_2018}.
However, previous local density approximation (LDA) calculations have revealed that this effect is very small and barely affects the band structure of the system~\cite{Wolf_PRL_2022}.
Thus we will ignore Rashba SOC. 
We model ($\sqrt{3}\times\sqrt{3}$)Sn/Si(111) using a single-band tight-binding Hamiltonian in the presence of on-site and extended Hubbard interactions on an isotropic triangular lattice, which is considered to be the simplest model for this material~\cite{Hansmann_PRL_2013,Li_NatComms_2013,Adler_2019}. 
The total Hamiltonian is given by
$\hat{\cal H}=\hat{\cal H}^{}_{0}+\hat{\cal H}^{}_{\rm int}$, in which the non-interacting part is expressed by
%
\begin{equation}
	\hat{\cal H}^{}_{0}=
	\sum_{\bk,\sigma} 
	\varepsilon^{}_{\bk}
	c^{\dagger}_{\bk\sigma}
	c^{}_{\bk\sigma} \, .
	\label{Eq:Ham_0}
\end{equation}
%
Here $c^{\dagger}_{\bk\sigma}$ ($c^{}_{\bk\sigma}$) creates (annihilates) an electron with momentum $\bk$ and spin $\sigma=\uparrow,\downarrow$. 
The kinetic energy $\varepsilon^{}_{\bk}$ includes hopping integrals up to fourth-nearest neighbors, and the tight-binding parameters are obtained from fitting the results of a first-principles local density approximation~\cite{Adler_2019}.
The non-interacting energy dispersion is then given by
\begin{align}
\begin{aligned}
	\varepsilon_{\bk}
	&=
	2t^{}_{1}
	\Big(
	\cos{k_x}+2\cos{\frac{k_x}{2}}\cos{\frac{\sqrt{3}k_y}{2}}
	\Big)
	\\
	&-2t^{}_{2}
	\Big(
	\cos{\sqrt{3}k_y}+2\cos{\frac{3 k_x}{2}}\cos{\frac{\sqrt{3}k_y}{2}}
	\Big)
	 \\
	&-2t^{}_{3}
	\Big( 
	\cos{2 k_x}+2\cos{k_x}\cos{\sqrt{3} k_y}
	\Big)
	\\
	&-4t^{}_{4}
	\Big(
	\cos{\frac{5k_x}{2}}\cos{\frac{\sqrt{3} k_y}{2}}
	 +
	 \cos{2 k_x}\cos{\sqrt{3}k_y}  \\
	&\qquad\quad
	 +
	 \cos{\frac{k_x}{2}}\cos{\frac{3\sqrt{3}k_y}{2}}
	\Big)
	-\mu,
\end{aligned}    
\end{align}
%
where the first neighbor hopping is set to $t_1=-52.7$ meV.
Relative to $t^{}_{1}$, the other hopping parameters are given by $t_2/t_1=-0.389$, $t_3/t_1=0.144$, and $t_4/t_1=-0.027$~\cite{Li_NatComms_2013,Wolf_PRL_2022}.
Besides, $\mu$ is the chemical potential, which varies with doping.
Throughout this paper, we set $\hbar=k^{}_{\rm B}=1$, and $T=0.03 t^{}_{1}$.
%
%

%
\begin{figure}[t]
	\begin{center}
		\vspace{0.10cm}
		\hspace{-0.3cm}
		\includegraphics[width=0.99 \linewidth]{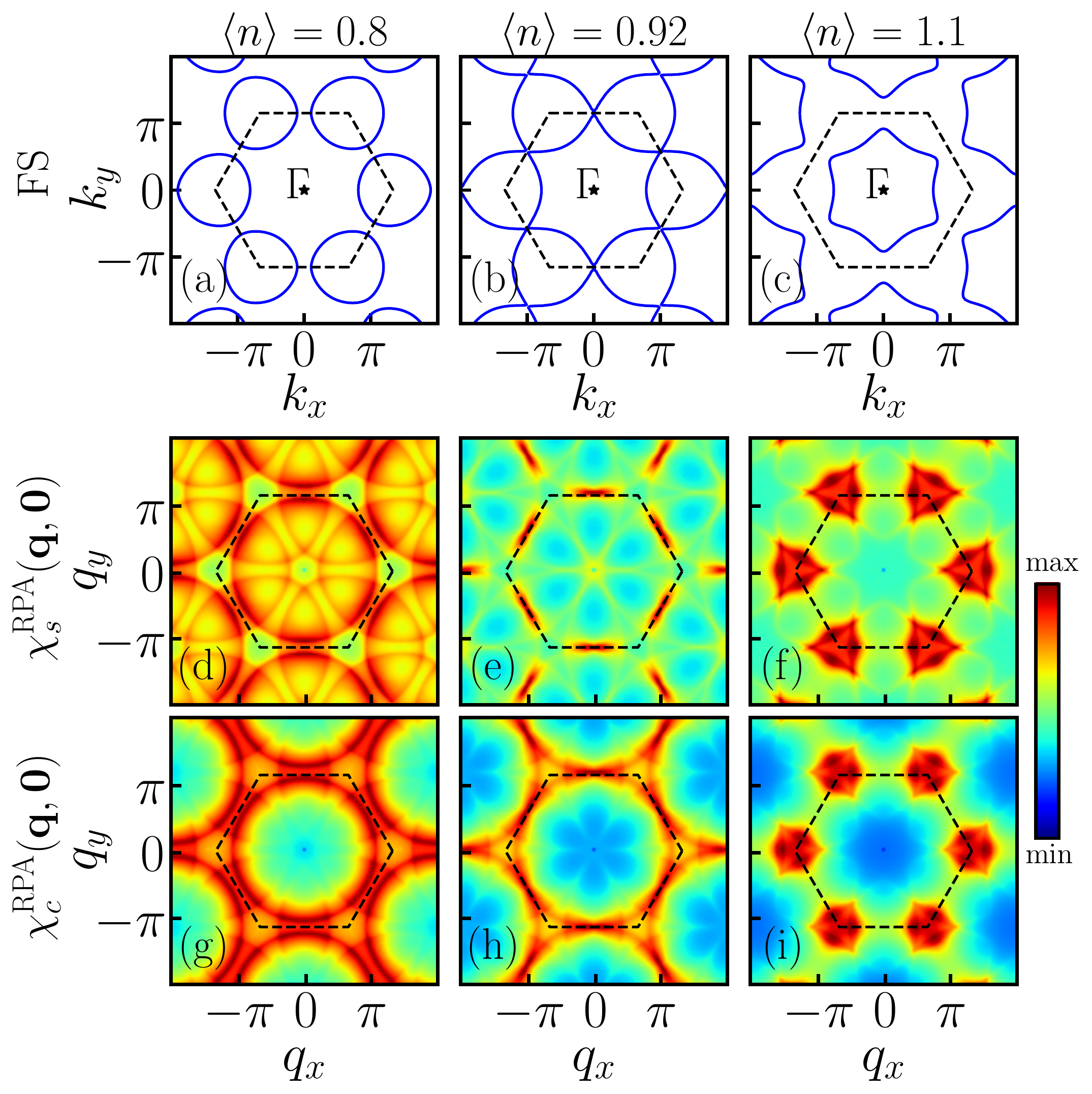}  \hspace{.0cm}
	\end{center}
	\vspace{-0.2cm}
	\caption{
		Fermi surface of ($\sqrt{3}\times\sqrt{3}$)-Sn/Si(111) for (a) $\langle n \rangle=0.8$, (b) $\langle n \rangle=0.92$, and (c) $\langle n \rangle=1.1$.
		(d)-(f): The corresponding RPA spin susceptibilities for $U^{}_{0}/t^{}_{1}=2.0$, and $V^{}_{0}/t^{}_{1}=1.0$, and
		(g)-(i) the RPA charge fluctuations for the levels of doping shown in (a)-(c).
		The black dashed lines mark the borders of the FBZ. 
	}
	\label{Fig:Fig_FS_Kappa_BZ}
\end{figure}
%

%
\begin{figure}[t]
	\begin{center}
		\vspace{0.0cm}
		\hspace{-1.05cm}
		\includegraphics[width=0.9 \linewidth]{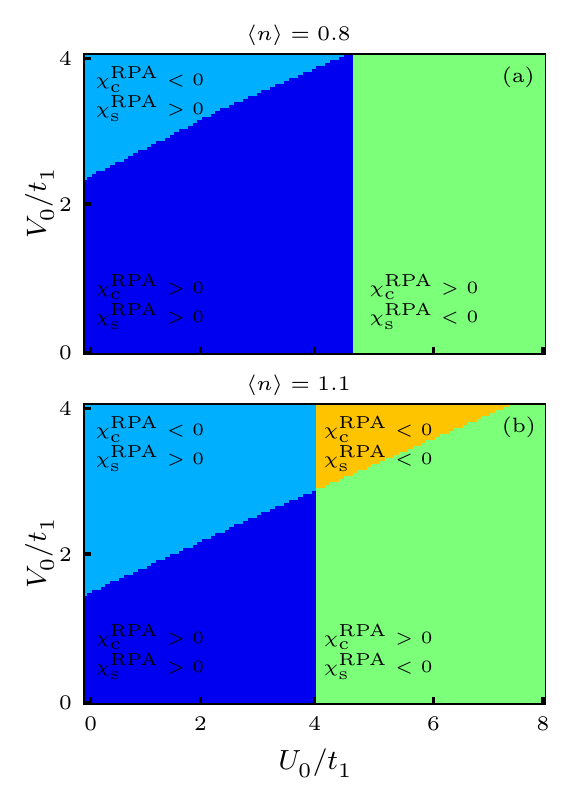}  \hspace{.0cm}
	\end{center}
	\vspace{-0.8cm}
	\caption{
		Phase diagram of charge and magnetic fluctuations in ($\sqrt{3}\times\sqrt{3}$)-Sn/Si(111) for different values of ($U^{}_{0}$,$V^{}_{0}$) based on the Stoner criterion for (a) $\langle n \rangle=0.8$ (hole-doped), and (b) $\langle n \rangle=1.1$ (electron-doped). 
		The dark blue areas denote the
		allowed phase space without any spin or charge long-range order.
		All remaining areas correspond to either long-range spin ($\chi_s^{\rm RPA}<0$) or charge ($\chi_c^{\rm RPA}<0$) orders, or both of them.
	}
	\label{Fig:Fig_CDW_SDW}
\end{figure}
%

%
\begin{figure*}[t]
	\begin{center}
		\vspace{0.0cm}
		\hspace{-1.05cm}
		\includegraphics[width=0.9 \linewidth]{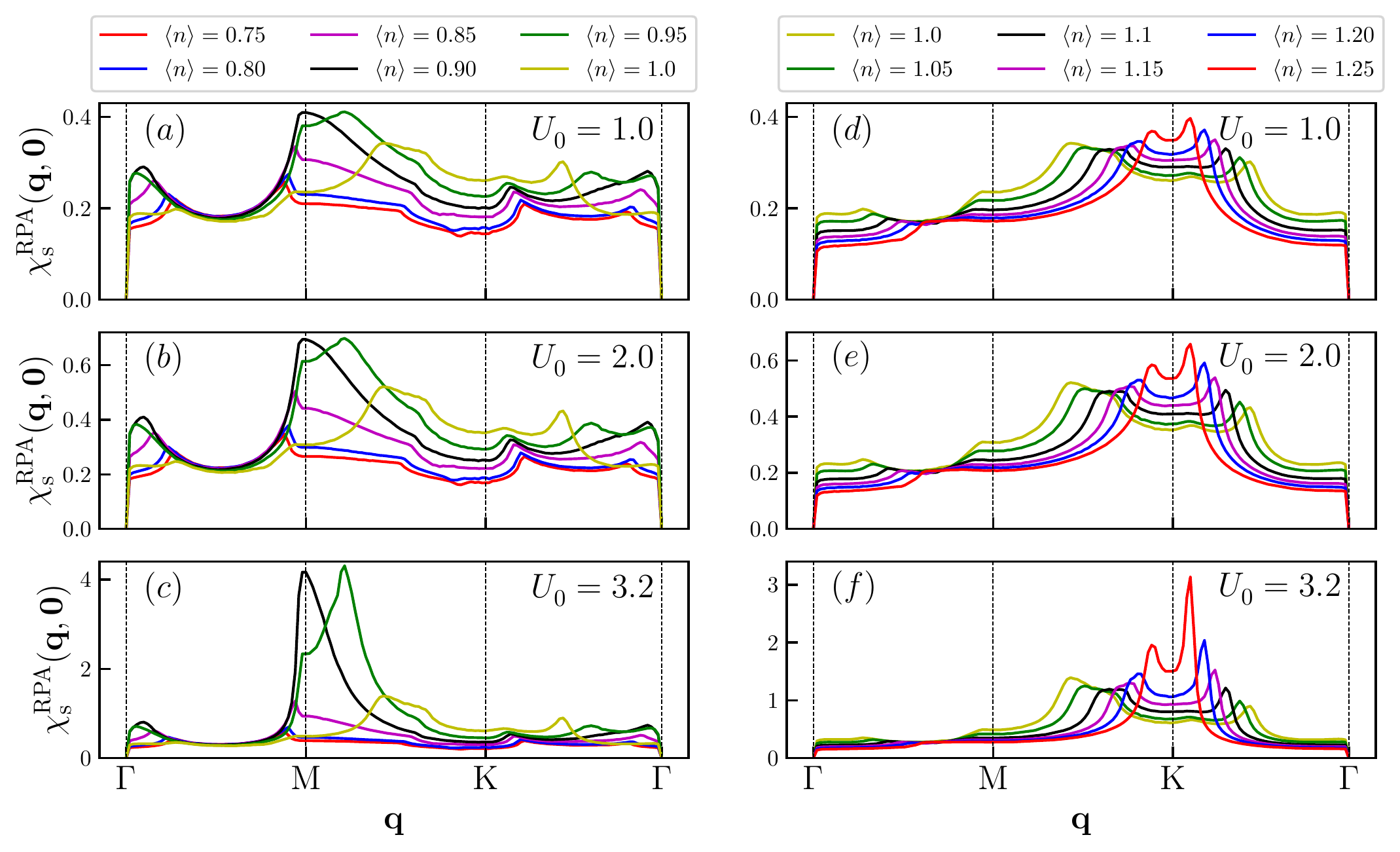}  \hspace{.0cm}
	\end{center}
	\vspace{-0.8cm}
	\caption{
		RPA spin susceptibility for hole-doping (left column) and electron-doping (right column) for different on-site Hubbard interactions $U_0$ and $V_0=0$ in the high symmetry $\Gamma$MK$\Gamma$ path of the BZ.
	}
	\label{Fig:Fig_RPA_Spin_Kappa}
\end{figure*}
%

%
\begin{figure*}[t]
	\begin{center}
		\vspace{0.0cm}
		\hspace{-1.05cm}
		\includegraphics[width=0.9 \linewidth]{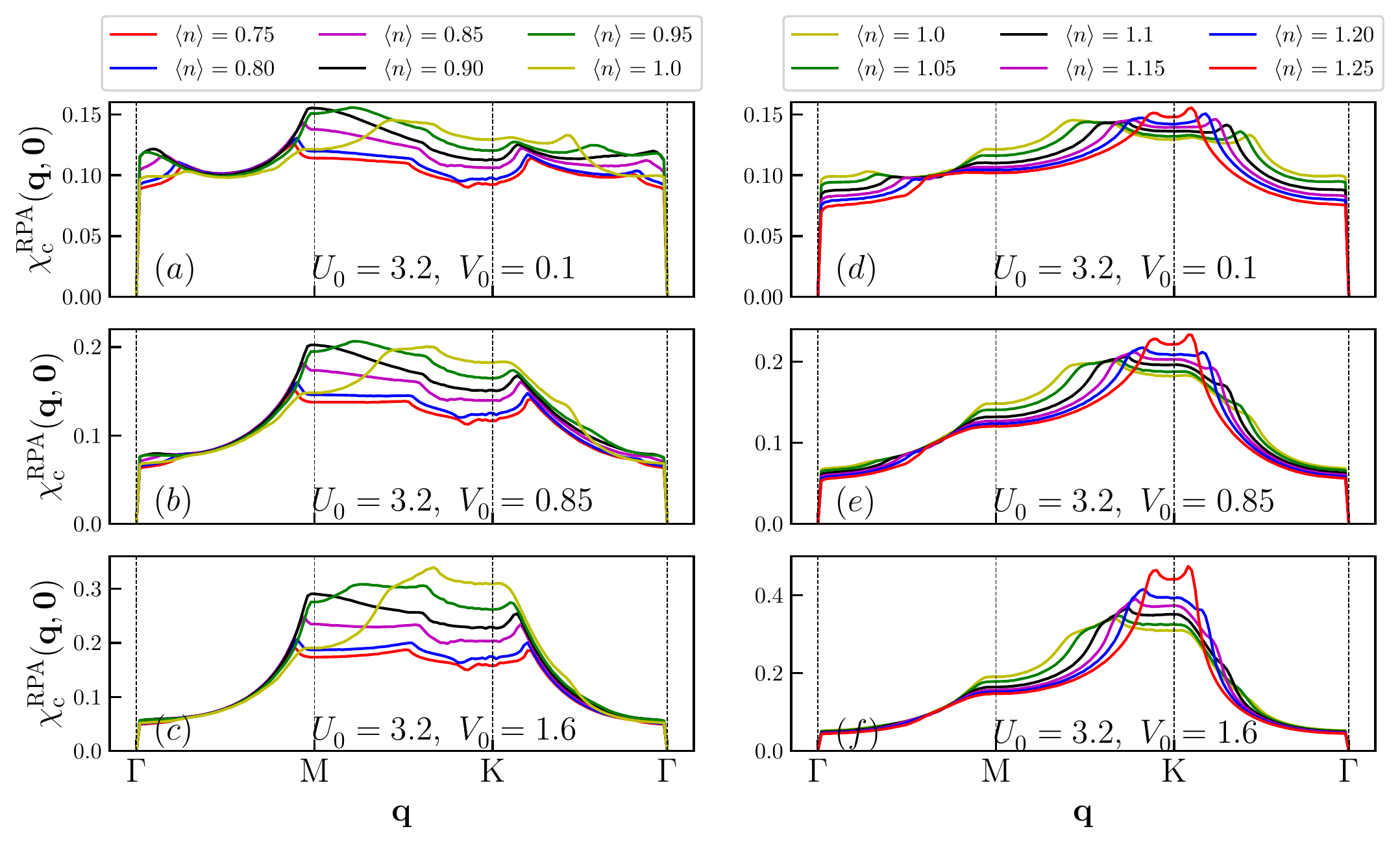}  \hspace{.0cm}
	\end{center}
	\vspace{-0.8cm}
	\caption{
		RPA charge susceptibilities for the hole-doped (left column) and electron-doped (right column) case for different Hubbard interactions strengths along the high symmetry $\Gamma$MK$\Gamma$ path of the BZ.
	}
	\label{Fig:Fig_RPA_Charge_Kappa}
\end{figure*}
%

%
\subsection{Charge and Spin Susceptibilities}
\label{SubSec:Charge_Spin_Kappa}
In the framework of linear response theory and using the Kubo formula, the spatial components of the bare susceptibility at temperature $T$ are given by~\cite{Biderang_PRB_2021}
%
\begin{align}
	\begin{aligned}
	\chi^{0}_{\alpha\beta}(\bq,{\mi}\omega^{}_{n})
	\!\!=\!\!
	-\frac{T}{N}
	\!\!\!\sum_{\bk,{\mi}\nu^{}_{m}}
	\!\!{\rm Tr}^{}_{\sigma}
	\Big[
	\hat{\sigma}^{}_{\alpha}
	&\hat{\cal G}^{0}_{}(\bk,{\mi}\nu^{}_{m})
	\\
	\times
	&\hat{\sigma}^{}_{\beta}
	\hat{\cal G}^{0}_{}(\bk+\bq,{\mi}\omega^{}_{n}+{\mi}\nu^{}_{m})
	\Big].
	\end{aligned}
\label{Eq:Bare_Kappa}	
\end{align}
%
In Eq.~(\ref{Eq:Bare_Kappa}), the summation is over momenta in the first Brillouin zone (FBZ).
The subscripts $\alpha,\beta\in\lbrace 0,x,y,z \rbrace$ refer to the spatial components of the bare susceptibility.
Here, $N$ denotes the number of grid points and is set to $N=18500$.
$\bk$ and $\bq$ correspond to electronic and bosonic momenta, and $\nu^{}_{m}=(2m+1)\pi T$ and $\omega^{}_{n}=2n\pi T$ to the fermionic and bosonic Matsubara frequencies, respectively.
The Pauli matrices are represented by the operator $\hat{\sigma}$.
Furthermore, 
$\hat{\cal G}^{0}_{}(\bk,{\mi}\nu^{}_{m})=
[{\mi}\nu^{}_{m} \hat{\mathbb{I}}
-
\hat{\cal H}^{}_{0}(\bk)]^{-1}_{}$
denotes the free electron Matsubara Green's function at momentum $\bk$ and frequency $\nu^{}_{m}$, in which $\hat{\mathbb{I}}$ denotes the $2\times 2$ identity matrix.
Due to the preserved SU($2$) symmetry, the non-diagonal elements of the bare susceptibility tensor are zero, and we can set  $\chi^{0}_{\alpha\beta}=\chi^{0}_{\alpha}\delta^{}_{\alpha,\beta}$.
%
%
%
Performing the summation over the fermionic Matsubara frequencies, the final form of the bare susceptibility is given by the following Lindhard function
%
%
\begin{equation}
	\chi^{0}_{}(\bq,{\mi}\omega^{}_{n})=
	\frac{1}{2N}
	\sum_{\bk} 
	\frac{
		n^{}_{\rm FD}(\varepsilon^{}_{\bk})
		-
		n^{}_{\rm FD}(\varepsilon^{}_{\bk+\bq})}
	{
		{\mi}\omega^{}_{n}
		-
		\varepsilon^{}_{\bk}
		+
		\varepsilon^{}_{\bk+\bq}
	},
	\label{Eq:Kappa_0_Lindhard}
\end{equation}
%
with $n^{}_{\rm FD}(\varepsilon^{}_{\bk})=[1+\exp(n^{}_{\rm F}(\varepsilon^{}_{\bk})/T)]^{-1}_{}$ being the Fermi-Dirac distribution function.
Within RPA, the $4\times 4$ matrix of susceptibilities dressed by the Hubbard interactions is given by
%
%
\begin{equation}
	\hat{\chi}^{\rm RPA}_{}(\bq,{\mi}\omega^{}_{n})=
	\frac
	{
	\hat{\chi}^{0}{}(\bq,{\mi}\omega^{}_{n})
	}
	{
	\hat{\mathbf{I}}
	-
	\hat{\cal V}(\bq)
	\hat{\chi}^{0}{}(\bq,{\mi}\omega^{}_{n})
	},
	\label{Eq:RPA_Kappa}
\end{equation}
%
where $\hat{\chi}^{0}{}(\bq,{\mi}\omega^{}_{n})=\sum_{\alpha\beta} \chi^{0}{}(\bq,{\mi}\omega^{}_{n})\delta^{}_{\alpha,\beta}$ and $\hat{\mathbf{I}}$ represent the bare susceptibility and the $4\times 4$ identity matrix, respectively.
Besides, $\hat{\cal V}(\bq)$ is the matrix of bare interactions that will be described in the next section.
Similar to the bare susceptibility, the matrix of RPA susceptibilities remains diagonal.
Using the analytical continuation ${\mi}\omega^{}_{n}\rightarrow\omega+{\mi}0^{+}_{}$, we can obtain the retarded RPA susceptibilities.
Finally, we are allowed to decompose the RPA susceptibility into separate charge and spin channels to investigate the charge and spin channels independently.
A simple calculation shows that the RPA charge $\chi^{\rm RPA}_{\rm c}$ and spin $\chi^{\rm RPA}_{\rm s}$ susceptibilities are given by
%
%
\begin{align}
		\chi^{\rm RPA}_{\rm s}(\bq,\omega)
		&=
		\frac
		{
			\chi^{0}{}(\bq,\omega)
		}
		{
			1
			-
			U^{}_{0}
			\chi^{0}{}(\bq,\omega)
		},
		\label{Eq:RPA_Spin_Kappa}	
		\\
		\chi^{\rm RPA}_{\rm c}(\bq,\omega)
		&=
		\frac
		{
			\chi^{0}{}(\bq,\omega)
		}
		{
			1
			+
			V(\bq)
			\chi^{0}{}(\bq,\omega)
		}.
		\label{Eq:RPA_Charge_Kappa}
\end{align}
%
The variables $U^{}_{0}$ and $V(\bq)$ refer to the Coulomb interactions and will be fully defined in the following section.
To investigate the dynamic effects and the role of frequency on the spin and charge fluctuations of the system, we introduce the dynamic spin and charge susceptibilities by 
$\chi^{\rm dyn}_{\rm s}(\bq,\omega)
=
{\rm Im}[\chi^{\rm RPA}_{\rm s}(\bq,\omega)]$,
and
$\chi^{\rm dyn}_{\rm c}(\bq,\omega)
=
{\rm Im}[\chi^{\rm RPA}_{\rm c}(\bq,\omega)]$, respectively.
Using these quantities, one can also define the spin and charge structure factors 
%
%
\begin{align}
	S(\bq,\omega)
	&=
	\frac
	{\chi^{\rm dyn}_{\rm s}(\bq,\omega)}
	{1-\exp(-|\omega|/T)},
	\label{Eq:Spin_Structure_Factor}
	\\	
	C(\bq,\omega)
	&=
	\frac
	{\chi^{\rm dyn}_{\rm c}(\bq,\omega)}
	{1-\exp(-|\omega|/T)}.	
	\label{Eq:Charge_Structure_Factor}
\end{align}
%

%
In addition, to gain a better understanding about the charge and spin orders in the system, we also calculate the charge density-charge density and spin density-spin density two-point correlation functions.
The former and latter ones are obtained by performing a
Fourier transformation of the charge and spin susceptibilities, respectively. 
At frequency $\omega$, the charge density-charge density correlation function is expressed as
%
\begin{equation}
	\langle n(0) n(\br) \rangle^{}_{\omega}=
	\frac{1}{4N}
	\sum_{\bq}
	\chi^{\rm RPA}_{\rm c}(\bq,\omega)
	\exp({\mi}\bq\cdot\br).	
	\label{Eq:Charge_Charge_Correlation}
\end{equation}
%
Furthermore, because of the preserved SU($2$) spin rotational symmetry, the in-plane and out-of-plane components of the spin density-spin density correlation functions are the same and given by
%
\begin{equation}
	\langle S(0) S(\br) \rangle^{}_{\omega}=
	\frac{1}{4N}
	\sum_{\bq}
	\chi^{\rm RPA}_{\rm s}(\bq,\omega)
	\exp({\mi}\bq\cdot\br).	
	\label{Eq:Spin_Spin_Correlation}
\end{equation}
%

\subsection{Effective Interaction}
\label{SubSec:Eff_Int}
The interacting part of the Hamiltonian contains both  on-site and nearest-neighbor terms.
In real space, one can write
\begin{equation}
	\hat{\cal H}^{}_{\rm int}
	=
	\frac{U^{}_{0}}{2}
	\sum_{i\sigma}
	c^{\dagger}_{i\sigma}    	
	c^{\dagger}_{i\bar{\sigma}}    
	c^{}_{i\bar{\sigma}}
	c^{}_{i\sigma}
	+
	\frac{V^{}_{0}}{2}
	\sum_{\langle ij\rangle\sigma\sigma'}
	c^{\dagger}_{i\sigma}    	
	c^{\dagger}_{j\sigma'}    
	c^{}_{j\sigma'}
	c^{}_{i\sigma},	
\end{equation}
%
with $U^{}_{0}$, and $V^{}_{0}$ as the amplitudes of the on-site and extended Hubbard interactions, respectively. Furthermore, $\langle ij \rangle$ denotes nearest-neighbors and $\bar{\sigma}=-\sigma$.
It has been suggested that to describe the experimental data, one should consider $1/3\!\!\leq\!\!V^{}_{0}/U^{}_{0}\!\!\leq\!\!1/2$~\cite{Hansmann_PRL_2013,Adler_2019}.
Performing a Fourier transform using $c^{\dagger}_{i\sigma}=\frac{1}{\sqrt{N}}\sum_{\bk}c^{\dagger}_{\bk\sigma}\exp({\mi}\bk\cdot\br)$, the interaction can be rewritten in momentum space as
%
\begin{equation}
	\hat{\cal H}^{}_{\rm int}
	=
	\!\!\!\sum^{}_{\bk\bk'\bq}
	\sum^{}_{\sigma\sigma'}
	{\cal U}(\bq)
	c^{\dagger}_{\bk+\bq\sigma}    
	c^{\dagger}_{\bk'-\bq\sigma'}
	c^{}_{\bk'\sigma'}	    
	c^{}_{\bk\sigma},
\end{equation}
%
with 
%
\begin{equation}
	{\cal U}(\bq)
	=
	\Big[
	2U^{}_{0}
	(1-\delta^{}_{\sigma,\sigma'})
	+
	2V^{}_{0}
	\Big(
	\cos q_x
	+
	2
	\cos \frac{q_x}{2}
	\cos \frac{\sqrt{3}q_y}{2}
	\Big)
	\Big].
\end{equation}
%
%
Defining the density operator
%
\begin{equation}
	\hat{\rho}^{}_{\bq\alpha}
	=
	\!\!\!\sum_{\bk,\sigma\sigma'}
	c^{\dagger}_{\bk+\bq\sigma}    
	\hat{\sigma}^{\alpha}_{\sigma\sigma'}
	c^{}_{\bk\sigma'},
\end{equation}
%
the interaction term can be rearranged as
%
\begin{equation}
	\hat{\cal H}^{}_{\rm int}
	=
	\sum_{\bq,\alpha\beta}
	\hat{\rho}^{}_{\bq\alpha}
	\hat{\cal V}^{}_{\alpha\beta}(\bq)
	\hat{\rho}^{}_{-\bq\beta},
\end{equation}
%
in which $\alpha=0$ and $\alpha\neq 0$ denote the charge and spin density operators, respectively.
The matrix of the bare interactions is given by
$\hat{\cal V}(\bq)={\rm diag}[V(\bq),-U^{}_{0},-U^{}_{0},-U^{}_{0}]$,
with
%
\begin{equation}
	V(\bq)=
	2U^{}_{0}
	+
	2V^{}_{0}
	\Big(
	\cos q_x
	+
	2
	\cos \frac{q_x}{2}
	\cos \frac{\sqrt{3}q_y}{2}
	\Big),
\end{equation}
%
and $w\in\lbrace x,y,z \rbrace$ runs over the spatial components of interaction matrix.
In the RPA framework, the effective (dressed) interaction is derived by a Dyson's equation
%
%
\begin{equation}
	\hat{\cal V}^{\rm RPA}_{}(\bq,\omega)=
	\frac
	{
		\hat{\mathbf{I}}
	}
	{
		\hat{\mathbf{I}}
		-
		\hat{\cal V}(\bq)
		\hat{\chi}^{0}{}(\bq,\omega)
	}
	\hat{\cal V}(\bq).
	\label{Eq:RPA_Int}
\end{equation}
%
It is obvious that $\hat{\cal V}^{\rm RPA}_{}(\bq,\omega)$ is a diagonal $4\times 4$ matrix, whose element with $\alpha=\beta=0$ is a pseudo-scalar  showing the effective interaction in the charge channel.
Therefore, one can write ${\cal V}^{}_{\rm c}(\bq,\omega)=\hat{\cal V}^{\rm RPA}_{}(\bq,\omega)\Big|^{}_{\alpha=\beta=0}$.
In the spin channel, the effective interaction is given by  ${\cal V}^{}_{\rm s}(\bq,\omega)={\rm Tr}[\hat{\cal V}^{\rm RPA}_{}(\bq,\omega)]\Big|^{}_{\alpha,\beta\neq 0}$.
Thus, we can define the total effective interaction as the sum of the charge and spin channels 
%
%
\begin{equation}
	{\cal V}^{}_{\rm eff}(\bq,\omega)
	=
	{\cal V}^{}_{\rm c}(\bq,\omega)
	+
	{\cal V}^{}_{\rm s}(\bq,\omega).
	\label{Eq:Final_Eff_Int}
\end{equation}
%
%
%

\begin{figure}[t]
	\begin{center}
		\vspace{0.0cm}
		\hspace{-0.7 cm}
		\includegraphics[width=1.07 \linewidth]{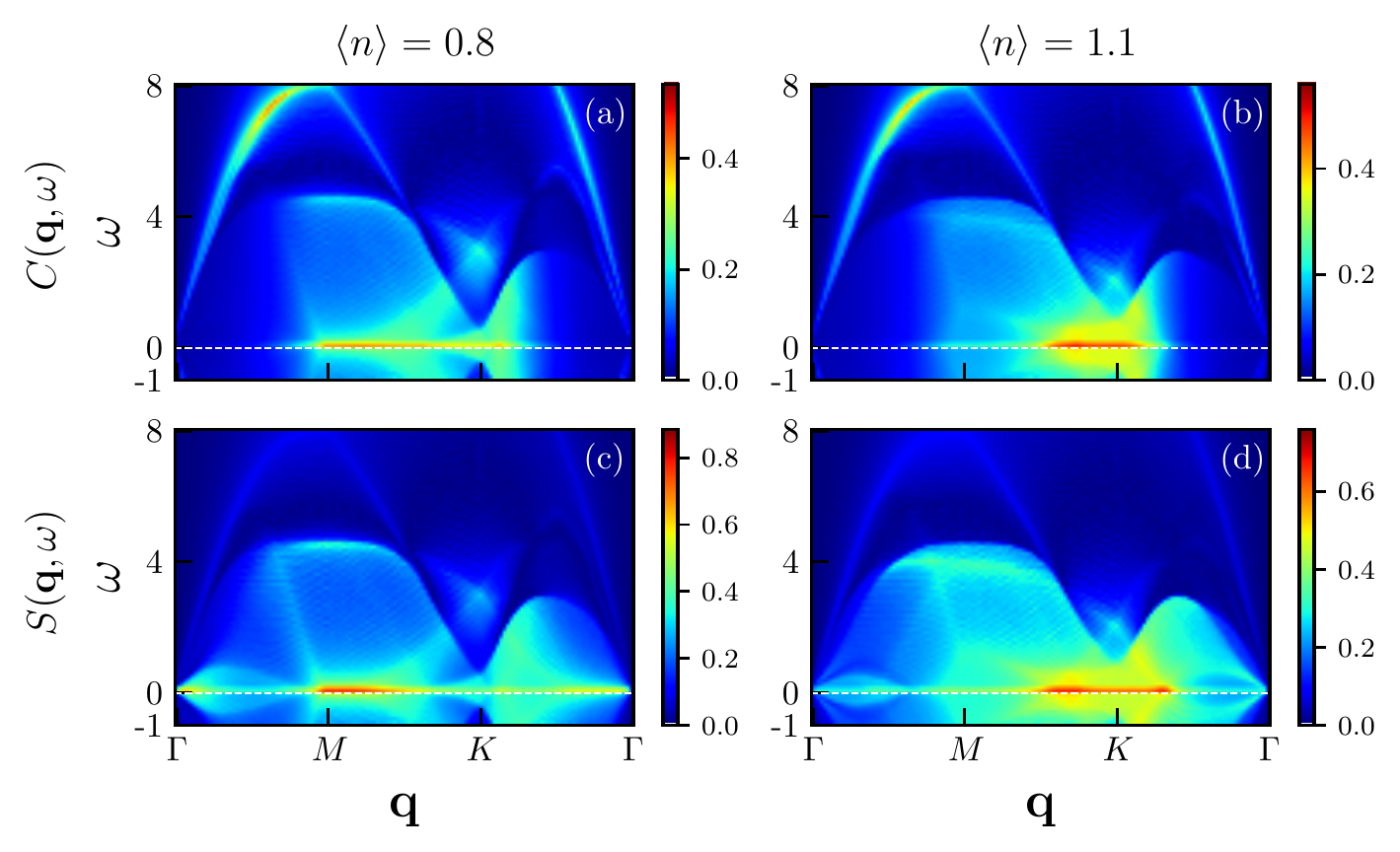}  \hspace{.0cm}
	\end{center}
	\vspace{-0.8cm}
	\caption{
		Charge structure factor for (a) $\langle n \rangle=0.8$, and (b) $\langle n \rangle=1.1$ with $U^{}_{0}/t^{}_{1}=2.0$, and $V^{}_{0}/t^{}_{1}=1.0$.
		(c), (d) show the corresponding spin structure factors.
	}
	\label{Fig:Structures}
\end{figure}

%
The results for the charge and spin structure factors shown in Fig.~\ref{Fig:Structures} reveal that the most significant contributions occur close to the Fermi surface.
Therefore, we will focus on the static susceptibilities and effective interactions, i.e.~$\omega=0$.
%
%

%
%

%
\subsection{Superconducting Instability}
\label{SubSec:SC}
Now we proceed to determine the symmetry of superconducting gap function. 
As the first step, we prefer to decompose the effective interaction into singlet and triplet channels, which are even and odd functions of momentum, respectively, and given by
%
\begin{equation}
	{\Gamma}^{\rm sing/trip}_{\rm eff}(\bk,\bk')
	=
	\frac{1}{2}
	\Big[
	{\cal V}^{}_{\rm eff}(\bk-\bk',0)
	\pm
	{\cal V}^{}_{\rm eff}(\bk+\bk',0)
	\Big].
	\label{Eq:Singlet_Triplet_Eff_Int}	
\end{equation}
%
%
Within BCS theory of superconductivity, the superconducting gap function can now be calculated self-consistently using~\cite{Sigrist_Rev_Mod_Phys_1991}
%
%
\begin{equation}
	\Delta^{}_{\bk}
	=
	-\frac{1}{N}
	\sum_{\bk'}
	{\Gamma}^{\rm sing/trip}_{\rm eff}(\bk,\bk')
	\frac
	{\Delta^{}_{\bk'}}
	{2E^{}_{\bk'}}
	\tanh
	\Big(
	\frac{E^{}_{\bk'}}{2T}
	\Big).
	\label{Eq:BCS_GAp_Function}
\end{equation}
%
In this equation, $\Delta^{}_{\bk}$ denotes the superconducting gap function and $E^{}_{\bk}=\sqrt{\varepsilon^{2}_{\bk}+\Delta^{2}_{\bk}}$ the energy of the superconducting quasiparticles.
Near the critical temperature $T^{}_{C}$, the BCS gap equation can be linearized and rewritten as an eigenvalue problem, whose dimensionless eigenvalues $\lambda$ carry the required information about the dominant gap function and the critical temperature $T^{}_{C}\propto \exp{(-1/\lambda)}$.
In the weak coupling limit for every individual angular momentum $l$, the dimensionless coefficient is given by~\cite{Greco_PRL_2018,Biderang_PRB_2022}
%
%
\begin{equation}
	\lambda^{}_{l}
	=
	-\frac
	{
	\int^{}_{\rm FS}
	\frac{dk}{v^{}_{F}(k)}
	\int^{}_{\rm FS}
	\frac{dk'}{v^{}_{F}(k')}
	\phi^{}_{l}(k)
	{\Gamma}^{\rm sing/trip}_{\rm eff}(\bk,\bk')
	\phi^{}_{l}(k')
	}
	{
		2\pi^2
	\int^{}_{\rm FS}
	\frac{dk'}{v^{}_{F}(k')}
	[\phi^{}_{l}(k')]^{2}_{}
	}.
	\label{Eq:Lambda_Function}
\end{equation}
%
In this equation, the momenta $k$ and $k'$ are restricted to the Fermi surface.
Moreover, $v^{}_{\rm F}(\bk)=|\boldsymbol{\nabla} E(\bk)|$ is the Fermi velocity.
Here, $\phi^{}_{l}(k)$ describes the momentum dependence of each allowed superconducting pairing, which are listed in Table.~\ref{Tab:SC_gap}.
The angular momentum corresponding to the largest eigenvalue $\lambda$ determines the dominant pairing symmetry of the system.

%
\begin{table}[ht]
	\centering
	\caption{Character table of the superconducting gap functions for different allowed angular momenta of the point group $D^{}_6$~\cite{Rachel_PRB_2018}.
	Note: Since the contributions of higher angular momenta are negligible, we refrain to report them.
	}
	\vspace{0.3 cm}
	\scalebox{1.1}{
		\begin{tabular}{c c c c} 
			\hline
			\hline
			$l$& Irrep. & Symmetry & $\phi^{l}_{\bk}$ \\ [1ex] 
			\hline
			0 & $A^{}_{1}$ & $s$-wave & $1$
			\\	
			0 & $A^{}_{1}$ & ext.$s$-wave & $\cos k_x + 2 \cos \frac{k_x}{2} \cos \frac{\sqrt{3} k_y}{2}$
			\\			
			1 & $E^{}_{1}$ & $p^{}_x$-wave & $\sin k_x + 2\sin \frac{k_x}{2} \cos \frac{\sqrt{3} k_y}{2}$ 
			\\
			1 & $E^{}_{1}$ & $p^{}_{y}$-wave & $ \cos \frac{k_x}{2} \sin \frac{\sqrt{3} k_y}{2}$
			\\
			2 & $E^{}_{2}$ & $d^{}_{x^2-y^2}$-wave & $\cos k_x - 2 \cos \frac{k_x}{2} \cos \frac{\sqrt{3} k_y}{2}$
			\\
			2 & $E^{}_{2}$ & $d^{}_{xy}$-wave & $ \sin \frac{k_x}{2} \sin \frac{\sqrt{3} k_y}{2}$ 
			\\
			3 & $B^{}_{1}$ & $f^{}_{x(x^2-3y^2)}$-wave & $\sin k_x - 2\sin \frac{k_x}{2} \cos \frac{\sqrt{3} k_y}{2}$
			\\
			\hline
			\hline
	\end{tabular}}
	\label{Tab:SC_gap}
\end{table}
%

%
In superconductors, measurement of the temperature dependence of the Knight shift at the resonance frequency $\delta \omega^{}_{s}$ is considered a very powerful experimental tool to determine the superconducting state of the system~\cite{Anderson_PRL_1959,Schriffer_PRL_1959,Ferrel_PRL_2959,Kadanov_PRL_1959,Cooper_PRL_1962,Fulde_PR_1965}.
Furthermore, the technique provides a reliable way to distinguish between the even and odd-parity Cooper pairings as well as the chiral and helical solutions~\cite{Frigeri_NJP_2004,Romer_PRB_2021}.
This experiment can be done using the nuclear magnetic resonance (NMR) technique and allows to find the spin structure of the Cooper pairs.
The evaluation of the ratio $\delta\omega^{}_{\rm sc}/\delta\omega^{}_{\rm n}={\rm Re}[\chi^{\rm sc}_{}(\bq=0,\omega=0)]/{\rm Re}[\chi^{0}_{}(\bq=0,\omega=0)]$ is a direct measure of the behaviour of the spin susceptibility of the
superconducting state w.r.t.~the normal ($\rm n$) state.
Here, $\chi^{0}_{}$ is the spin-susceptibility of the normal state given by Eq.~(\ref{Eq:Kappa_0_Lindhard}).
Besides, $\chi^{\rm sc}_{}$ denotes the superconducting spin susceptibility.
Further details about the calculation of the superconducting spin susceptibility and of the Knight shift are given in App.~\ref{App:Knight_Shift}.
%

%
\section{Results}
\label{Sec:Results}
Next, we discuss in detail the results for ($\sqrt{3}\times\sqrt{3}$)Sn/Si(111) obtained by using the methods described in the previous section. 
\subsection{Electronic band Structure and Susceptibilities}
\label{SubSec:Res_Band_Kappa}
Fig.~\ref{Fig:Fig_Geometry_Band}(c) shows the tight-binding band structure of the system.
This simple band structure originates from the dangling-bond surface states, which have a two-fold spin degeneracy.
At half-filling, every upward dangling bond has exactly one electron.
Moreover, a saddle point in the band structure exists at the M point resulting in a van Hove singularity in the density of states (Fig.~\ref{Fig:Fig_Geometry_Band}(d)).
Figs.~\ref{Fig:Fig_FS_Kappa_BZ}(a)-(c) show the evolution of the Fermi surface with respect to filling for $\langle n \rangle=0.8$, $0.92$, and $1.1$, respectively.
For the hole-doped case with $\langle n \rangle=0.8$, the Fermi surface has electron-like pockets centered around the K points in the BZ.
At $\langle n \rangle=0.92$, the contours of the Fermi surface touch the borders of the BZ at the M points.
This filling marks a Lifshitz transition, which results in a van Hove singularity in the DOS and a changing of the topology of the Fermi surface.
At $\langle n \rangle=1.1$, the topology of the Fermi surface has been changed into hole-like textures, with pockets which are centered around the $\Gamma$ points.
Figs.~\ref{Fig:Fig_FS_Kappa_BZ}(d)-(i) show the RPA spin and charge susceptibilities at $U^{}_{0}/t^{}_{1}=2.0$, and $V^{}_{0}/t^{}_{1}=1.0$ for different levels of filling.
As discussed earlier, the spin and charge channels can be decoupled to study the fluctuations and instabilities in each individual channel separately.
%
%
We should emphasize again that due to the preserved spin rotational SU$(2)$ symmetry, the longitudinal and transverse components of the spin susceptibility are the same.
%
%
From these figures it is clear that for the shown fillings and interaction strengths, the system hosts incommensurate magnetic and charge fluctuations.
In Fig.~\ref{Fig:Fig_CDW_SDW}, we plot the charge and magnetic phase diagram for two different levels of filling based on the Stoner criterion.
The idea here is that at critical values of $U^{}_{0}$ ($V^{}_{0}$) in the spin (charge) channel, the determinant of the denominator of Eq.~(\ref{Eq:RPA_Kappa}) will vanish, indicating a diverging susceptibility and thus a transition into an order phase~\cite{Ghadimi_PRB_2019,Schordi_PRB_2020}.
For the spin and charge channels, this instability is known as a spin-density wave (SDW) or a charge-density wave (CDW), respectively.
%
%
Apart from a region (dark blue in Fig.~\ref{Fig:Fig_CDW_SDW}) where both RPA susceptibilities are positive and neither spin nor charge order is established, the system also shows SDW (light green), CDW (light blue), and both types of long-range order (yellow) based on the Stoner condition.
We note that SDW and CDW order only coexist in the electron-doped case.
%
On a qualitative level, we also see that the Stoner criterion predicts that the system orders already for smaller interation values in the electron-doped than in the hole-doped case.
%

%
Fig.~\ref{Fig:Fig_RPA_Spin_Kappa} provides a more detailed analysis of the spin fluctuations in the system as a function of filling and the magnitude of the on-site Hubbard interaction for $V^{}_{0}=0$.
Eq.~(\ref{Eq:RPA_Spin_Kappa}) shows that the RPA spin susceptibility is only affected by the on-site Hubbard interaction which is why we can set the strength of the nearest-neighbor Coulomb repulsion to zero in this case.
For the hole-doped case, shown in Figs.~\ref{Fig:Fig_RPA_Spin_Kappa}(a)-(c), the peak of the RPA spin susceptibilities remains near the M point in the BZ.
Decreasing the concentration of holes in the compound leads to an increase of the intensity of spin fluctuations near the M point but the fluctuations are incommensurate.
For $\langle n \rangle=0.9$, the peak of the spin susceptibility reaches the M point implying commensurate magnetic fluctuation.
This trend remains valid until the van Hove singularity at $\langle n \rangle =0.92$ is approached. At this doping level, the position of the peak starts shifting towards the K point.
Right at half-filling, the maximum lies somewhere between the M and the K points.
%
%
%
In the electron doped case, the intensity of the peak near the K point increases with the doping level, see Figs.~\ref{Fig:Fig_RPA_Spin_Kappa} (d)-(f). The spin fluctuations, however, remain incommensurate with the lattice. 
%

%
\begin{figure}[t]
	\begin{center}
		\vspace{0.10cm}
		\hspace{-0.3cm}
		\includegraphics[width=0.99 \linewidth]{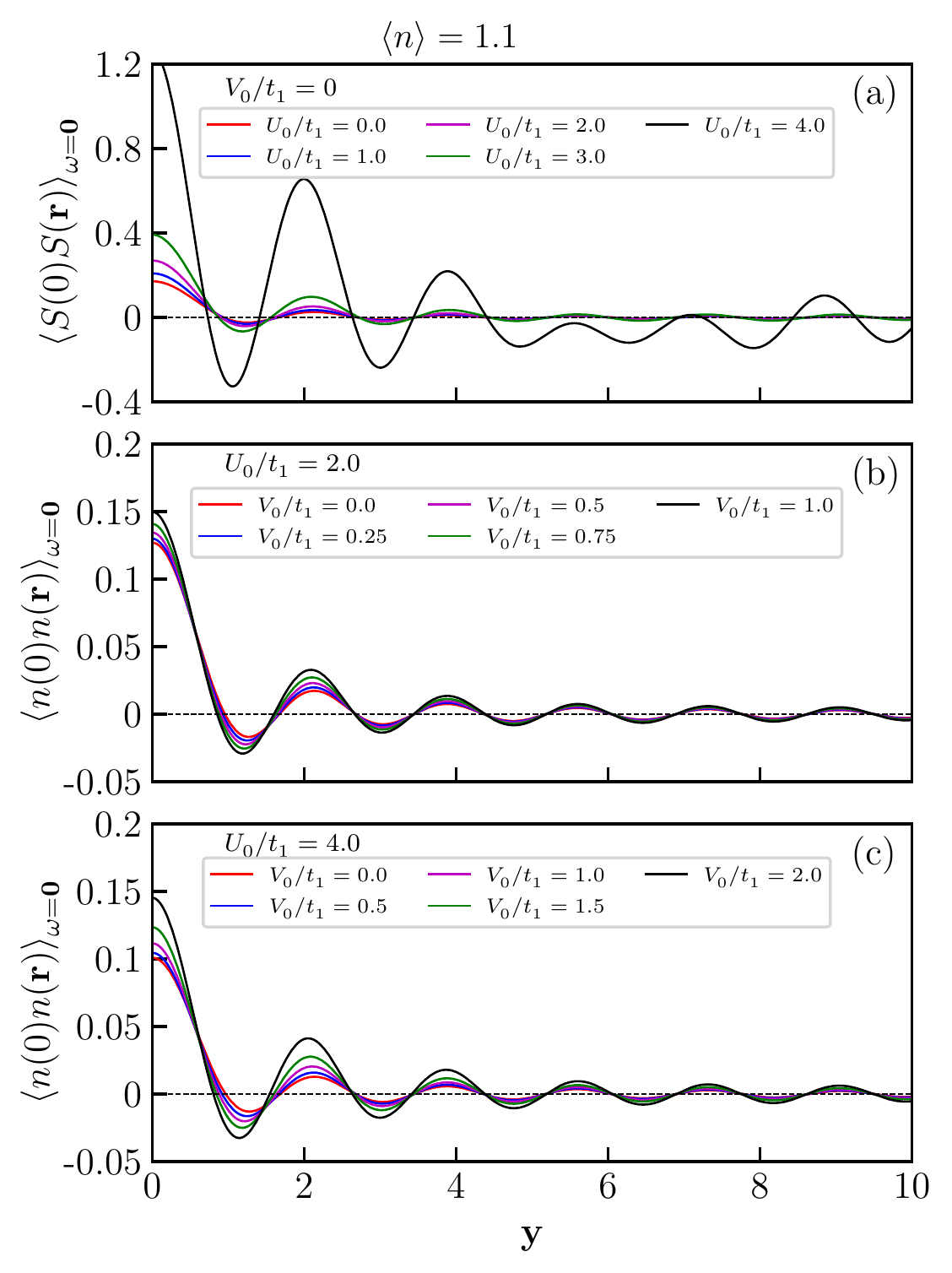}  \hspace{.0cm}
	\end{center}
	\vspace{-0.2cm}
	\caption{
		(a) Spin density-spin density correlation function for ($\sqrt{3}\times\sqrt{3}$)-Sn/Si(111) in $y$-direction for varous on-site Hubbard interactions at $\langle n \rangle=1.1$.
		(b), (c) Charge density-charge density correlation function for different extended Hubbard interactions at $\langle n \rangle=1.1$ for $U^{}_{0}/t^{}_{1}=2.0$, and $U^{}_{0}/t^{}_{1}=4.0$, respectively.
	}
	\label{Fig:Orders}
\end{figure}
%

%
\begin{figure}[t]
	\begin{center}
		\vspace{-0.1cm}
		\hspace{-0.9cm}
		\includegraphics[width=0.99 \linewidth]{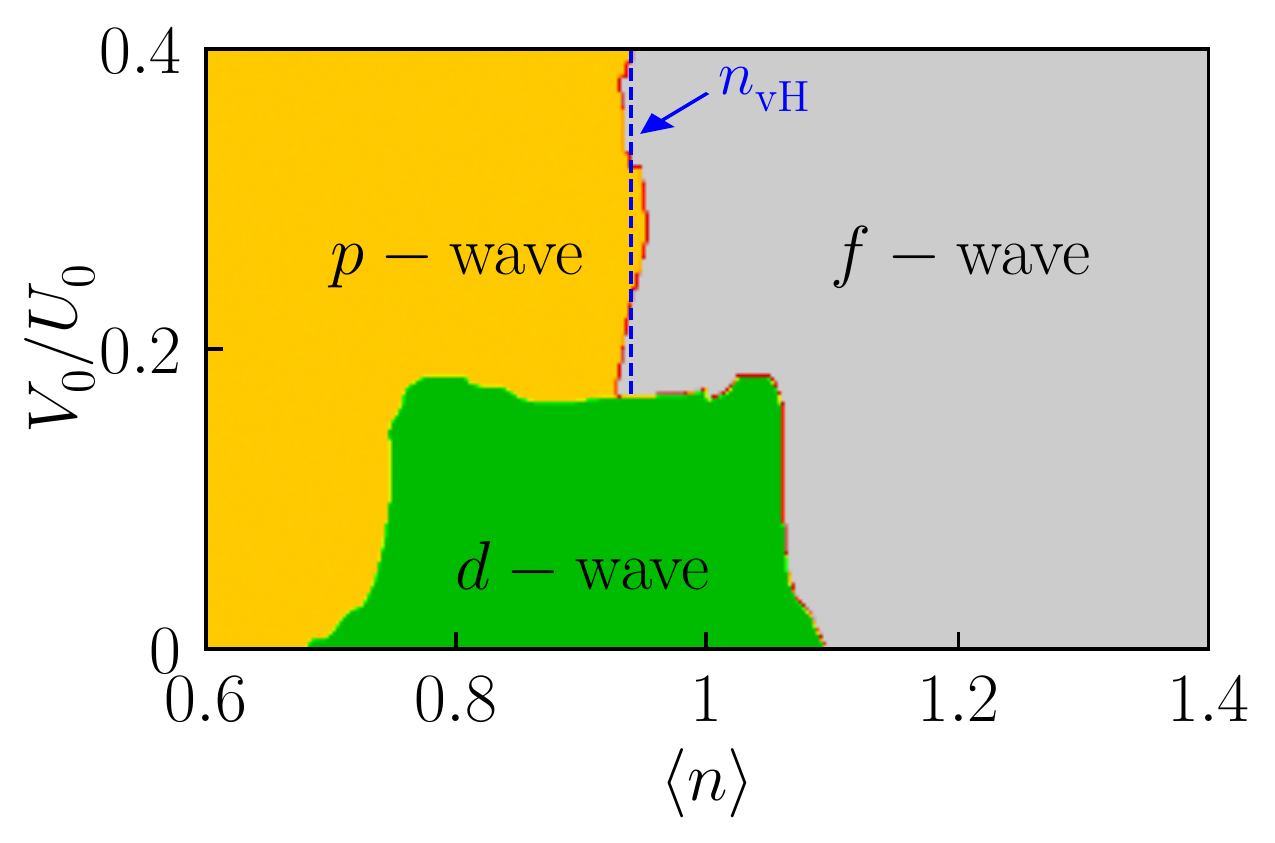}  \hspace{.0cm}
	\end{center}
	\vspace{-0.8cm}
	\caption{
		Superconducting phase diagram of ($\sqrt{3}\times\sqrt{3}$)-Sn/Si(111) as a function of band filling and interaction ratio $V^{}_{0}/U^{}_{0}$ obtained within weak coupling theory.
		The blue dashed-line represents the position of the van Hove singularity at $\langle n \rangle=0.98$.
		%
		For $V^{}_{0}/U^{}_{0}\lesssim 0.2$ and near half-filling, the superconducting order parameter has chiral $d^{}_{x^2-y^2}\pm{\mi}d^{}_{xy}$-wave texture, a non-trivial topology with Majorana fermions at the edges of the sample.
		For $V^{}_{0}/U^{}_{0}\gtrsim 0.2$, spin-triplet pairing dominates.
		%
		%
	 For $\langle n \rangle<0.98$, the topological chiral $p^{}_{x}\pm{\mi}p^{}_{y}$-wave spin-triplet superconductivity appears while for
		$\langle n \rangle>0.98$, the triplet odd-parity $f$-wave pairing is realized.
		%
		%
	}
	\label{Fig:Sc}
\end{figure}
%

%
\begin{figure}[t]
	\begin{center}
		\vspace{0.0cm}
		\hspace{-1.05cm}
		\includegraphics[width=0.99 \linewidth]{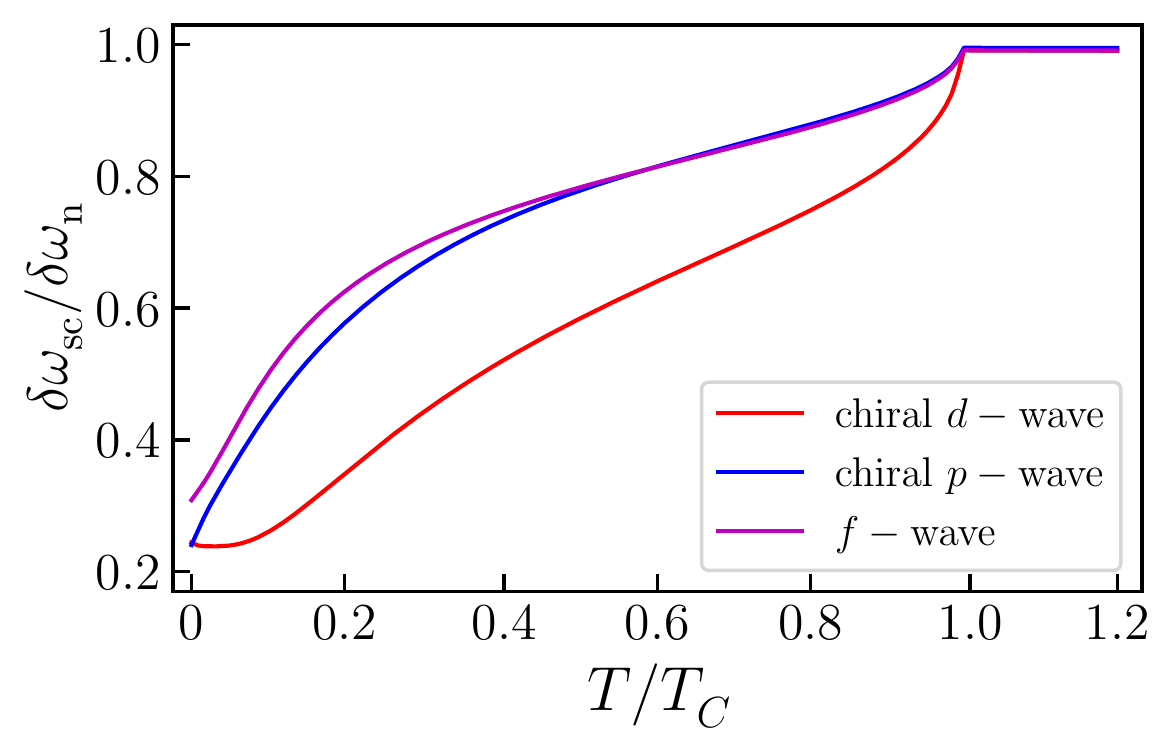}  \hspace{.0cm}
	\end{center}
	\vspace{-0.8cm}
	\caption{
		Knight shift in ($\sqrt{3}\times\sqrt{3}$)-Sn/Si(111) for the different pairings depicted in Fig.~\ref{Fig:Sc}.
		The blue, red, and magenta solid lines show the chiral $p$-, chiral $d$-, and $f$-wave superconductivity at $\langle n \rangle=0.65$, $0.9$, and $1.2$, respectively.
		Here, we set $k^{}_{\rm B}T^{}_{C}\approx 0.4$ meV and the maximum amplitude of the gap to $\Delta^{}_{\rm max}=0.3t^{}_{1}$.
	}
	\label{Fig:Knight_Temp}
\end{figure}
%

%
The charge fluctuations of the system within RPA as a function of filling and interaction strength are shown in  Fig.~\ref{Fig:Fig_RPA_Charge_Kappa}.
Here, the left and right columns correspond to the hole- and electron-doped cases, respectively.
Since the amplitude of the on-site Coulomb interaction does not have a significant impact on the structure of the charge susceptibility (see Eq.~(\ref{Eq:RPA_Charge_Kappa})), we keep $U^{}_{0}/t^{}_{1}=3.2$ fixed.
In the hole-doped case (Figs.~\ref{Fig:Fig_RPA_Charge_Kappa}(a)-(c)) the maxima lie in $\Gamma$-M path, very close to the M point of the BZ.
Increasing the concentration of holes leads to a decline of the magnitude of charge fluctuations similar to the spin fluctuations discussed above.
No long-range order is established for $V^{}_{0}/U^{}_{0}\leq1/2$.
Figs.~\ref{Fig:Fig_RPA_Charge_Kappa}(d)-(f) show that in the electron-doped case the maxima of the RPA charge fluctuations shift to regions around the K point when the concentration of charge carriers is increased.
%
%
Moreover, increasing the interaction towards $V^{}_{0}/t_1\sim 2$ drives the system into a CDW ordered state.
This result is consistent with the phase diagram shown in Fig.~\ref{Fig:Fig_CDW_SDW}.
Figs.~\ref{Fig:Structures}(a), (b) show the charge structure factors for $U^{}_{0}/t^{}_{1}=2.0$, $V^{}_{0}/t^{}_{1}=1.0$ at $\langle n \rangle=0.8$, and $\langle n \rangle=1.1$, respectively.
The corresponding spin structure factors $S(\bq,\omega)$ are depicted in Figs.~\ref{Fig:Structures} (c) and (d).
The plots clearly show that the most significant contributions come from frequencies very close to the Fermi surface at $\omega=0$.
%
%
Consistent with the previous results, we find again that for $\langle n \rangle=0.8$, the most important charge and spin effects occur near the M point in the BZ while for $\langle n \rangle=1.1$, the dominant fluctuations move to the K point.

\subsection{Charge and Spin Orders}
\label{SubSec:Res_Orders}
To summarize our findings about the spin and charge fluctuations in the system, we investigate the zero-frequency two-point spin density-spin density and charge density-charge density correlation functions along the y-direction.
Fig.~\ref{Fig:Orders}(a) depicts the zero-frequency spin density-spin density correlation function $\langle S(0) S(\br) \rangle^{}_{\omega=0}$ versus $U^{}_{0}/t^{}_{1}$ at $\langle n \rangle=1.1$.
Based on what we have learned from Fig.~\ref{Fig:Fig_CDW_SDW}(b), we expect that at filling $\langle n \rangle=1.1$ the system enters a SDW ordered phase at $U^{}_{0}/t^{}_{1}\approx 4.0$. 
From Fig.~\ref{Fig:Orders}(a) it is clear that for $U^{}_{0}/t^{}_{1}\!\!<\!\!4.0$, the system shows only short-range antiferromagnetic fluctuations.
However, for $U^{}_{0}/t^{}_{1}\geq 4.0$, a phase transition occurs and the system shows slightly incommensurate but long-range antiferromagnetic fluctuations with small amplitude.
The charge density correlations $\langle n(0) n(\br) \rangle^{}_{\omega=0}$ are shown in Figs.~\ref{Fig:Orders} (b), and (c) for $U^{}_{0}/t^{}_{1}=2.0$, and $4.0$, respectively.
For $V^{}_{0}/U^{}_{0}\leq 1/2$, the system does not show any long-range CDW order. 
However, for $V^{}_{0}/U^{}_{0}> 1/2$ incommensurate long-range CDW order appears.

\subsection{Superconducting Instability}
\label{SubSec:Res_SC}
One of the main results of our study is the superconducting phase diagram of ($\sqrt{3}\times\sqrt{3}$)Sn/Si(111), shown in Fig.~\ref{Fig:Sc}, obtained in weak coupling theory. It shows the dominant superconducting instabilities as a function of filling and the ratio of nearest to on-site Coulomb interactions $V^{}_{0}/U^{}_{0}$.
Around half-filling and for small $V^{}_{0}/U^{}_{0}$, $d$-wave pairing is the most favourable.
In this case, the spin-singlet $d$-wave superconductivity has the chiral structure of $d^{}_{x^2-y^2}\!\pm{\mi}d^{}_{xy}$ with spontaneously broken time-reversal symmetry (TRS).
For hole-doping far away from half-filling, the effect of hexagonal warping of the Fermi surface leads to spin-triplet pairing~\cite{Thomale_PRL_2013}.
In addition, spin-triplet superconductivity is also realized for all fillings for ratios $V^{}_{0}/U^{}_{0}\gtrsim 0.2$.
Here, $p$-wave pairing is realized for fillings below the van-Hove singularity and odd-parity spin-triplet $f$-wave dominates for fillings above the van-Hove singularity. The transition between the two right at the singularity highlights the important role of the band structure for the superconducting gap function.
Fig.~\ref{Fig:Sc} also shows that the nearest-neighbor interaction does play an important role and can turn spin-singlet to spin-triplet superconductivity. 
As shown in Figs.~\ref{Fig:Orders}(b), (c), increasing the ratio $V^{}_{0}/U^{}_{0}$ amplifies the strength of charge fluctuations in the system which then facilitates the formation of spin-triplet superconductivity.
The warping of the Fermi surface due to the presence of larger distance hopping processes is another factor which strengthens spin-triplet pairing.
For future works, it would be interesting to investigate the interplay between the warping of the Fermi surface due to longer-range hoppings and the effect of non-local Coulomb interactions in more detail.
Our results obtained within RPA are in very good agreement with Ref.~\cite{Wolf_PRL_2022} based on functional renormalization group, both qualitatively and quantitatively.
Finally, in Fig.~\ref{Fig:Knight_Temp} we show the temperature dependence of the Knight shift for the three different superconducting order parameters shown in Fig.~\ref{Fig:Sc}.
Because of the preserved SU(2) spin-rotational symmetry, the measured Knight shifts remain the same for both in-plane and out-of-plane magnetic fields.
While the results for the chiral $p$-wave and the $f$-wave order are qualitatively similar, the chiral $d$-wave shows a much stronger temperature dependence near the transition, setting it apart from the other two cases. 
In the absence of SOC, one would expect that the Knight shift
of the even-parity superconductors should be completely 
suppressed in all spin channels for $T\rightarrow 0$ in an exponential/linear fashion for a full/nodal gap function~\cite{Yosid_PR_1958,Romer_PRL_2019}.
However, it seems that the chiral characteristic of superconductivity breaks down this fact and pushes up the Knight shift of an even-parity chiral d-wave system.
For the odd-parity pairing, there is always a residual Knight shift even at zero temperature.
This result is consistent with previous results for the case of chiral p-wave superconductivity in Sr$^{}_{2}$RuO$^{}_{4}$~\cite{Ishida_Nature_1998}.
%

%
%

\section{Conclusions}
We have theoretically investigated the pairing mechanism and the gap symmetry for the recently discovered superconductivity in doped ($\sqrt{3}\times\sqrt{3}$)Sn/Si(111) using the RPA in the framework of linear response theory.
Following previous works, we modeled the system using a triangular tight-binding Hamiltonian excluding the antisymmetric Rashba SOC, which has been shown to have a negligible effect on the electronic band structure.
Our study has been focused on the effects of filling and the ratio of nearest-neigbor versus on-site Coulomb interaction on the charge and magnetic fluctuations as well as on the superconducting state.
To study the magnetic and charge fluctuations, we have calculated the RPA spin and charge susceptibilities.
Using the Stoner criterion, we have identified the charge and magnetic phase diagram of the system and found the transition lines between short-range and long-range order.
Calculations of the dynamical charge and spin susceptibilities and of the structure factors show that the most significant contributions come from the vicinity of the Fermi surface at $\omega=0$.
However, where these main contributions are situated in the BZ, strongly depends on the level of filling.
For example, at $\langle n \rangle=1.1$, the most dominant charge and spin fluctuations happen around the K point in BZ at $\omega=0$, leading to incommensurate charge and magnetic fluctuations.
To further study the nature of the charge and magnetic orders in the system, we have calculated the zero-frequency two-point charge density-charge density and spin density-spin density correlation functions, respectively.
We have found that in the regime of $V^{}_{0}/U^{}_{0}\leq1/2$, the system does not show any long-range charge or spin order.
%
%

%
Based on a linearized BCS gap equation, we have obtained the phase diagram of the leading superconducting instability with respect to  filling and the interaction ratio $V^{}_{0}/U^{}_{0}$.
We have found that around half-filling and for smaller values of $V^{}_{0}/U^{}_{0}$, the superconducting ground state has a chiral $d$-wave texture.
Away from half-filling, two different superconducting states are realized.
In the hole doped case, the system shows a chiral $p$-wave symmetry. 
This state is topologically non-trivial and belongs
to the C class of topological superconductivity characterized by a $\mathbb{Z}$ invariant.
For electron-doping, on the other hand, the odd-parity spin triplet $f$-wave dominates.
Our study shows that charge fluctuations together with the hexagonal warping of the FS are the most likely mechanisms favouring spin-triplet Cooper pairing in  ($\sqrt{3}\times\sqrt{3}$)Sn/Si(111).
Finally, we obtained the temperature dependence of the Knight shift for these three superconducting instabilities. The latter results might be helpful in experimental investigations of the symmetry of the superconducting gap function.
\\

\section*{ACKNOWLEDGMENTS}
We are grateful to R. Thomale, A. Akbari, A. R${\o}$mer, M. Malakhov, S. A. Jafari, and M. N. Najafi for fruitful discussions. 
M.H.Z. was supported by Grant No.
G546139, research deputy of Qom Uiversity of Technology.
J.S. acknowledges support by the Natural Sciences and Engineering Research Council (NSERC, Canada) and by the
Deutsche Forschungsgemeinschaft (DFG) via Research Unit
FOR 2316.
%

\appendix

\section{Calculation of Knight Shift}
\label{App:Knight_Shift}
Here, we discuss a method to calculate the Knight shift as the ratio of the spin susceptibility in the superconducting state versus that in the normal phase.
Within linear response theory and for the external magnetic field along $\alpha\in\lbrace x,y,z\rbrace$, the Knight shift can be formulated
in terms of the real part of the static spin-resolved susceptibility
at $\bq=0$.
The superconducting spin susceptibility $\chi^{\rm sc}_{\alpha}$ can be expressed as~\cite{Frigeri_NJP_2004}
%
%
\begin{align}
	\begin{aligned}
		\chi^{\rm sc}_{\alpha\beta}(\bq,{\mi}\omega^{}_{n})
		\!\!=\!\!
		\frac{-T}{4N}
		\!\!\!
		\sum_{\bk,{\mi}\nu^{}_m}
		\!\!\!
		{\rm Tr}^{}_{\sigma}
		\Big[
		&\check{\sigma}^{}_{\alpha}
		\check{G}(\bk,{\mi}\nu^{}_m)
		\check{\sigma}^{}_{\alpha}
		\check{G}(\bk\!+\!\bq,{\mi}\nu^{}_m\!\!+{\mi}\omega^{}_{n})	
		\Big].
	\end{aligned}
	\label{Eq:SC_Kappa}
\end{align}
%
%
In this equation, $\check{\sigma}^{}_{\alpha}$ is a $4\times 4$ Pauli matrix in the particle-hole symmetric Nambu space defined by
%
%
\begin{equation}
	\check{\sigma}^{}_{\alpha}
	=
	\begin{bmatrix}
		\hat{\sigma}^{}_{\alpha} & 0
		\\
		0 & -\hat{\sigma}^{\intercal}_{\alpha}
	\end{bmatrix}.
\end{equation}
%
Moreover, the matrix of the Matsubara Green's function $\check{G}(\bk,{\mi}\nu^{}_{m})$ is given by
%
\begin{equation}
	\check{G}(\bk,{\mi}\nu^{}_{m})
	=
	\begin{bmatrix}
		\hat{G}(\bk,{\mi}\nu^{}_{m}) & \hat{F}(\bk,{\mi}\nu^{}_{m})
		\\
		\hat{F}^{\dagger}_{}(\bk,{\mi}\nu^{}_{m}) 
		& 
		-\hat{G}^{\intercal}_{}(-\bk,-{\mi}\nu^{}_{m})
	\end{bmatrix}.
\end{equation}
%
In this equation, $\hat{G}(\bk,{\mi}\nu^{}_m)$ and $\hat{F}(\bk,{\mi}\nu^{}_m)$ are the normal and anomalous Green's functions of a superconducting state and given by~\cite{Mahan_ManyBody_2000}
%
%
\begin{align}	
	\hat{G}(\bk,{\mi}\nu^{}_m)
	&=
	\Big[
	\frac{u^{2}_{\bk}}{{\mi}\nu^{}_m-E^{}_{\bk}}
	+
	\frac{v^{2}_{\bk}}{{\mi}\nu^{}_m+E^{}_{\bk}}
	\Big]\hat{\sigma}^{}_{0},
	\label{Eq:Normal_Green}
	\\		
	\hat{F}(\bk,{\mi}\nu^{}_m)
	&=
	-u^{}_{\bk}v^{}_{\bk}
	\Big[
	\frac{1}{{\mi}\nu^{}_m-E^{}_{\bk}}
	-
	\frac{1}{{\mi}\nu^{}_m+E^{}_{\bk}}
	\Big]\hat{\sigma}^{}_{0},
	\label{Eq:Anomalous_Green}
\end{align}
%
where the coherence factors $u^{}_{\bk}$ and $v^{}_{\bk}$ are
%
%
\begin{equation}
	\begin{Bmatrix}
		u^{}_{\bk}
		\\
		v^{}_{\bk}
	\end{Bmatrix}
	=
	\sqrt{
		\frac{1}{2}
		\Big(
		1
		\pm
		\frac
		{\varepsilon^{}_{\bk}}
		{E^{}_{\bk}}
		\Big)
	}.
	\label{Eq:Coherence_factors}
\end{equation}
%
Preserved spin rotational SU(2) symmetry requires that $\chi^{\rm sc}_{\alpha}=\chi^{\rm sc}_{}$.
Therefore, the real part of the static superconducting spin susceptibility is defined by
%
%
\begin{equation}
	{\rm Re}[\chi^{\rm sc}_{}(\bq,\omega=0)]
	=
	-\frac{1}{2N}\sum_{\bk}
	\Big[
	{\cal A}^{}_{}(\bk,\bq)
	-
	{\cal B}^{}_{}(\bk,\bq)
	\Big],
	\label{Eq:SC_Kappa_Bare}
\end{equation}
%
where,
%
\begin{align}
	\begin{aligned}
		{\cal A}^{}_{}(\bk,\bq)
		&-
		{\cal B}^{}_{}(\bk,\bq)
		=
		\\
		&\Big[
		\frac
		{n^{}_{\rm FD}(E^{}_{\bk})-n^{}_{\rm FD}(E^{}_{\bk+\bq})}
		{E^{}_{\bk}-E^{}_{\bk+\bq}}
		\Big]
		\Big(
		u^{}_{\bk}u^{}_{\bk+\bq}
		-
		v^{}_{\bk}v^{}_{\bk+\bq}
		\Big)^{2}_{}
		\\
		+
		&\Big[
		\frac
		{n^{}_{\rm FD}(E^{}_{\bk})+n^{}_{\rm FD}(E^{}_{\bk+\bq})-1}
		{E^{}_{\bk}+E^{}_{\bk+\bq}}
		\Big]
		\Big(
		u^{}_{\bk}v^{}_{\bk+\bq}
		+
		v^{}_{\bk}u^{}_{\bk+\bq}
		\Big)^{2}_{}.
	\end{aligned}
	\label{Eq:A_minus_B}
\end{align}
%
It can be easily seen that in the limit $\bq\to 0$ one obtains
%
\begin{align}
	\begin{aligned}
	{\rm Re}[\chi^{\rm sc}_{}(0,0)]
	=
	-\frac{1}{2N}\sum_{\bk}
	\Big[
	&(u^{2}_{\bk}-v^{2}_{\bk})^{2}_{}
	\frac{\partial n^{}_{\rm FD}(E^{}_{\bk})}{\partial E^{}_{\bk}}
	\\
	&+2u^{2}_{\bk}v^{2}_{\bk}
	\frac{[2n^{}_{\rm FD}(E^{}_{\bk})-1]}{E^{}_{\bk}}
	\Big].
	\end{aligned}
	\label{Eq:Main_Local_Sc_Kappa}
\end{align}
%
%
%
Using the Fermi-Dirac distribution, Eq.~\eqref{Eq:Main_Local_Sc_Kappa} can be written explicitly as
%
\begin{equation}
	{\rm Re}[\chi^{\rm sc}_{}(0,0)]
	=
	\frac{1}{2N}\sum_{\bk}
	\Big[
	\frac{(u^{2}_{\bk}-v^{2}_{\bk})^{2}_{}}{4T\cosh^{2}_{}\frac{E^{}_{\bk}}{2T}}
	+
	\frac{2u^{2}_{\bk}v^{2}_{\bk}}{E^{}_{\bk}}
	\tanh\frac{E^{}_{\bk}}{2T}
	\Big].
	\label{Eq:SC_Kappa_Bare_Local}
\end{equation}
%
It should be noted that the temperature dependence of the superconducting gap magnitude within the BCS theory is modelled by~\cite{Romer_PRL_2019}
%
\begin{equation}
	\Delta^{}_{\bk}(T)
	=
	\Delta^{}_{\bk}
	\tanh\Big[
	1.76\sqrt{\frac{T^{}_{C}}{T}-1}
	\Big].
	\label{Eq:SC_gap_T_dependence}
\end{equation}
%

%
Using Eq.~(\ref{Eq:Kappa_0_Lindhard}), it can be shown that for $\omega=0$ and $\bq=0$, the real part of the non-superconducting spin susceptibility is given by
%
%
\begin{equation}
	{\rm Re}[\chi^{0}_{}(0,0)]
	=
	\frac{1}{2N}\sum_{\bk}
	\frac{1}{4T\cosh^{2}_{}\frac{\varepsilon^{}_{\bk}}{2T}}.
	\label{Eq:Normal_Kappa_Bare_Local}
\end{equation}
%
In the limit of $T\rightarrow T^{}_{C}$, Eq.~(\ref{Eq:SC_Kappa_Bare_Local}) reduces to Eq.~(\ref{Eq:Normal_Kappa_Bare_Local}), i.e., these two equations are consistent.
Now, we can easily calculate the Knight shift for every individual superconducting gap function using
\begin{equation}
	\frac{\delta\omega^{}_{\rm sc}}{\delta\omega^{}_{\rm n}}
	=
	\frac
	{{\rm Re}[\chi^{\rm sc}_{}(\bq=0,\omega=0)]}
	{{\rm Re}[\chi^{0}_{}(\bq=0,\omega=0)]}.
	\label{Eq:Final_Knight_Shift_Definition}
\end{equation}
%
Substituting Eqs.~(\ref{Eq:SC_Kappa_Bare_Local}), and~(\ref{Eq:Normal_Kappa_Bare_Local}) into Eq.~(\ref{Eq:Final_Knight_Shift_Definition}), one finds the final expression for the Knight shift which we have used in the manuscript.
%

\bibliography{References}
%

%
\end{document}